*Submitted to the Journal of Electron Spectroscopy*
# Depth-Resolved Composition and Electronic Structure of Buried Layers and Interfaces in a LaNiO$_3$/SrTiO$_3$ Superlattice from Soft- and Hard- X-ray Standing-Wave Angle-Resolved Photoemission

D. Eiteneer[1,2], G. K. Pálsson[1,2,&], S. Nemšák[1,2,3], A. X. Gray[1,2,4,@], A. M. Kaiser[1,2], J. Son[6,#], J. LeBeau[6,$], G. Conti[1,2], A. A. Greer[1,2,5], A. Keqi[1,2], A. Rattanachata[1,2], A. Y. Saw[1,2], A. Bostwick[7], E. Rotenberg[7], E. M. Gullikson[8], S. Ueda[9,13], K. Kobayashi[9,%], A. Janotti[6], C. G. Van de Walle[6], A. Blanca-Romero[11,12], R. Pentcheva[10,11], C. M. Schneider[3], S. Stemmer[6], C. S. Fadley[1,2]

[1]*Department of Physics, University of California, Davis, California 95616, USA*
[2]*Materials Sciences Division, Lawrence Berkeley National Laboratory, Berkeley, California 94720, USA*
[3]*Peter-Grünberg-Institut PGI-6, Forschungszentrum Julich, 52425 Julich, Germany*
[4]*Stanford Institute for Materials and Energy Sciences, SLAC National Accelerator Laboratory, Menlo Park, California 94025, USA*
[5]*Chemical Engineering and Materials Science Engineering, University of California, Davis, California 95616, USA*
[6]*Materials Department, University of California, Santa Barbara, California 93106, USA*
[7]*Advanced Light Source, Lawrence Berkeley National Laboratory, Berkeley, California 94720, USA*
[8]*Center for X-Ray Optics, Lawrence Berkeley National Laboratory, Berkeley, California 94720, USA*
[9]*NIMS Beamline at SPring-8, National Institute for Materials Science, Hyogo 679-5148, Japan*
[10]*Theoretical Physics, University of Duisburg-Essen and Center of Nanointegration (CENIDE), Duisburg, 47057 Germany*
[11]*Department of Earth and Environmental Sciences, Section Crystallography and Center of Nanoscience, University of Munich, D-80333 Munich, Germany*
[12]*Department of Chemistry, Thomas Young Centre, Imperial College London, London SW7 2AZ, United Kingdom*
[13]*Quantum Beam Unit, National Institute for Materials Science, Tsukuba 305-0047, Japan*
[&]*Present address: Department of Physics, Uppsala University, Uppsala, SE-751 20 Sweden*
[@]*Present address: Department of Physics, Temple University, Philadelphia, PA 19122, USA*
[#]*Present address: Department of Materials Science and Engineering, Pohang University of Science and Technology, Pohang, 790-784, Republic of Korea*
[$]*Present address: Department of Materials Science and Engineering, North Carolina State University, Raleigh, NC 27695 USA*
[%]*Present address: Japan Atomic Energy Agency, Hyogo 679-5148, Japan*

2**Abstract**

LaNiO$_3$ (LNO) is an intriguing member of the rare-earth nickelates in exhibiting a metal-insulator transition for a critical film thickness of about 4 unit cells [Son et al., Appl. Phys. Lett. 96, 062114 (2010)]; however, such thin films also show a transition to a metallic state in superlattices with SrTiO$_3$ (STO) [Son et al., Appl. Phys. Lett. 97, 202109 (2010)]. In order to better understand this transition, we have studied a strained LNO/STO superlattice with 10 repeats of [4 unit-cell LNO/3 unit-cell STO] grown on an (LaAlO$_3$)$_{0.3}$(Sr$_2$AlTaO$_6$)$_{0.7}$ substrate using soft x-ray standing-wave-excited angle-resolved photoemission (SWARPES), together with soft- and hard- x-ray photoemission measurements of core levels and densities-of-states valence spectra. The experimental results are compared with state-of-the-art density functional theory (DFT) calculations of band structures and densities of states. Using core-level rocking curves and x-ray optical modeling to assess the position of the standing wave, SWARPES measurements are carried out for various incidence angles and used to determine interface-specific changes in momentum-resolved electronic structure. We further show that the momentum-resolved behavior of the Ni 3d $e_g$ and $t_{2g}$ states near the Fermi level, as well as those at the bottom of the valence bands, is very similar to recently published SWARPES results for a related La$_{0.7}$Sr$_{0.3}$MnO$_3$/SrTiO$_3$ superlattice that was studied using the same technique (Gray et al., Europhysics Letters **104**, 17004 (2013)), which further validates this experimental approach and our conclusions. Our conclusions are also supported in several ways by comparison to DFT calculations for the parent materials and the superlattice, including layer-resolved density-of-states results.**I. Introduction**

Perovskite nickelates are examples of materials exhibiting Mott metal-insulator transitions that lie at the heart of correlated electron physics. LaNiO$_3$ (LNO), however, is at the boundary of the family of rare earth nickelates and remains a 3D Fermi liquid at all temperatures[1]. It has attracted considerable interest when a seemingly dimensionality-induced metal-insulator transition (MIT) was found in ultrathin films of LNO[2,3,4]. The Fermi surface of a thin film of LNO on SrTiO$_3$ (STO) was mapped out recently using angle-resolved photoemission (ARPES) by Eguchi and co-workers[1], who discovered a relatively flat Fermi surface with a correlation-related kink at 0.25 eV below the Fermi level. This was put into a theoretical perspective by Balents et al.[5], who showed, using a model Hamiltonian consisting of Ni 3d $e_g$ states only, that electron states with the correct symmetry and band dispersion could be reconstructed. This model was, however, unable to explain the MIT found in thin films and heterostructures in terms of confinement only and suggested a physical mechanism whereby the potential at the interface partially polarizes the *d* states so as to conduct poorly. Recently, our group has shown using standing-wave photoemission that in heterostructures of LNO and STO a suppression of near-Fermi states of LNO takes place, an observation which is indicative of a MIT close to the LNO/STO interfaces[6]. However, the thickness of the LNO (4 unit cells (u.c.)) at which this was observed corresponds to the four-layer case in the work of Balents[5], which predicted an enhancement of the density of states (DOS) at the Fermi level rather than a suppression. This superlattice is also a member of a family where enhanced conductivity was found by Stemmer et al.[7]. It is fair to say that there is no conclusion at the moment as to what constitutes the conditions whereby LNO becomes insulating or metallic in thin films and heterostructures. There is thus clearly a need for further experimental evidence to elucidate the electronic structure at the LNO/STO interface, in order to better understand the circumstances under which a MIT in thin LNO takes place.

Hwang et al.[8] investigated the structure of the same type of LNO/STO superlattice as in the present paper and concluded that the $NiO_6$ octahedral tilts in LNO originate from a requirement of maintaining connectivity across the STO interface. It is clear that the Ni $e_g$- and $t_{2g}$- derived states should be affected by these tilts and that such effects might be observable in further more depth-sensitive photoemission measurements on this system, including standing-wave excitation to provide enhanced depth resolution, and involving both soft x-ray (i.e. several hundred eV) and hard x-ray (several keV) excitation to vary the probing depth.

In pioneering theoretical studies of related oxide heterostructure systems, Chaloupka and Khaliullin[9] and Hansmann et al.[10] predicted that the LNO Fermi surface could be turned into one resembling a cuprate through heterostructuring with e.g. $LaAlO_3$, and the experimental realization of this prediction by Liu et al. for $LaAlO_3$[11] showed intriguing similarities between the two Fermi surfaces. These results thus raise the question of what exactly the LNO electronic structure looks like when sandwiched with $SrTiO_3$.

We have thus undertaken a detailed standing-wave photoemission and ARPES study of this system, making use of methodologies applied by Gray et al.[12,13] to $La_{0.7}Sr_{0.3}MnO_3/SrTiO_3$ (LSMO/STO), Kaiser et al.[6] to LNO/STO, and Nemšák et al.[14] to $GdTiO_3/SrTiO_3$ (GTO/STO).

This work makes use of a special computer program for modeling standing-wave effects in photoemission by Yang et al.[15]. As a specific point, we make direct comparison of the electronic structure of LSMO/STO as observed in SWARPES to that of LNO/STO and highlight the similarities between the $e_g$ and $t_{2g}$ states of the transition metal Mn and Ni components.

The standing-wave photoemission technique provides a unique capability for depth-resolving the electronic structure of the superlattice, particularly in its momentum-resolving form of standing-wave ARPES (SWARPES) using soft x-rays in the ca. 1 keV regime[6,12,13,14]. The main advantages of this technique are its non-destructiveness, the much larger effective attenuation lengths compared to the much more common vacuum-ultraviolet ARPES that is normally done below about 150 eV, and the enhanced depth resolution for buried interfaces via standing-wave excitation. The SWARPES method is based on the interference of the incoming and outgoing x-ray, which sets up a standing wave (SW) inside and outside of a thin film. The SW can be varied in depth by either changing the incidence angle or the photon energy. When the sample is a multilayer, a strong reflection at the first-order Bragg condition amplifies the SW and also dictates that the wavelength of SW intensity perpendicular to the surface will be very close to the multilayer period $d_{ML}$. Being able to tune the SW in depth by varying incidence angle, thus generating a rocking curve (RC), or photon energy, thus permits a local enhancement or suppression of the regions directly at the interfaces between two materials or in the central regions of the two materials. Further details on this method and the theoretical modeling of it can be found elsewhere[4,6,12,13,14,15].

## II. Experimental procedure

The superlattice was grown by rf magnetron sputtering deposition and consisted of 10 bilayers of LNO (4 u.c. = 15.6 Å) and STO (3 u.c. = 11.7 Å) grown on a $(LaAlO_3)_{0.3}(Sr_2AlTaO_6)_{0.7}$ (LSAT) substrate, with the sample and experimental configuration shown in Fig. 1. The top layer of STO was found in a prior study, and confirmed in this one, to be reduced in thickness to about 9 Å[6]. The surface is (001) oriented, and the superlattice is tensile strained by 0.78% in the plane of the film due to epitaxial growth on the substrate with experimental pseudocubic lattice parameters of 3.87 Å[8]. The SW periodicity is thus $d_{ML}$ = 11.7 Å + 15.6 Å = 27.3 Å (cf. Fig. 1).

The photoemission experiments have been carried out at two synchrotron radiation beamlines (BLs): BL 7.0.1 of the Advanced Light Source (ALS, soft x-rays, with photon energy set to 833.2 eV, just below the La $M_5$ absorption resonance) and BL15XU of SPring-8 (hard x-rays, photon energy set to a non-resonant 5953.4 eV), with complementary measurements at BL 9.3.1 of the ALS (hard x-rays at 2550, 3238, and 4000 eV) discussed elsewhere[16]. All of the measurements have been carried out without any kind of sample surface cleaning after exposure to atmosphere, with this being possible due to the relative stability of the topmost STO layer and its surface, and the higher photon energies, which permit probing below this surface. As measures of this, the inelastic mean free paths (IMFPs) in STO (LNO) for our two photon energies will be approximately 17 Å (14 Å) at 833.2 eV and 83 Å (69 Å) at 5953.4 eV, as estimated from the TPP-2M formula as included in the SESSA electron spectroscopy simulation program[17]. Beyond this, survey and individual core-level spectra were taken before and after each ARPES series to verify the lack of significant surface chemical alteration or radiation-induced damage to the sample.

The soft x-ray data should thus be most sensitive to the first STO/LNO bilayer, while the hard x-ray data should sample several STO/LNO bilayers. However, since the x-ray field penetrates through the whole stack, the SW modulation felt within the depth sensitivity of photoemission originates from interference of the x-rays throughout the whole structure, and will thus exhibit effects due to both Bragg reflection according to $n\lambda_x = 2d_{ML}sin\theta_{Bn}$, where $n$ is the order, $\lambda_x$ the x-ray wavelength, $d_{ML}$ the bilayer thickness and $\theta_B$ the Bragg angle, and Kiessig fringes due to reflection from the top and bottom of the entire multilayer stack, according to $m\lambda_x = 2D_{ML}sin\theta_{K,m}$, where $m$ is the order, $D_{ML}$ is the thickness of the entire multilayer, and $\theta_{K,m}$ is the $m^{th}$ order angle for constructive interference. The STO stability was checked with periodic measurements of Sr, Ti, O, and C relative intensities with soft or hard x-ray photoemission. The initial survey and core-level spectra, as well as rocking curve (RC) measurements, were carried out at room temperature, leading to valence-band spectra that should represent matrix-element weighted densities of states (MEWDOS)[13]. Momentum-resolved ARPES measurements were carried out at cryogenic temperatures to minimize the effects of phonon smearing and x-ray photoelectron diffraction on the angle-resolved data[13]; the temperature varied slightly, over 20-45 K, during our measurements, but this should have no effect on the results. The end station at BL 7.0.1 was equipped with a hemispherical electron analyzer (VG Scienta R4000), in a geometry with the angle between incident x-rays and the analyzer equal to 60° and an overall energy resolution of ~300 meV, and the end station at BL15XU also equipped with a VG Scienta

R4000, with the angle between incident x-rays and the outgoing electrons equal to 90°, and an overall energy resolution of 230 meV. The resolution and position of the zero binding energy were frequently measured using the Au 4*f* peak position and the Au Fermi edge of a reference sample. In the soft x-ray experiments, the incidence angle was varied from 13.0° to 17.6° to span the first-order Bragg angle of the 4 u.c. LNO/3 u.c. STO multilayer, which is expected to be at approximately $\sin^{-1}[\lambda_x/2d_{ML}] = 15.8°$, where $\lambda_x$ is the x-ray wavelength of 14.87 Å.

### III. Theoretical calculations

Two types of calculations were performed, one to determine the band offset at the LNO/STO interface, and another to determine the detailed electronic structure in the multilayer.

**A. Bulk band structures and band offset**: First-principles calculations were performed for STO and LNO in order to determine the band alignment at the interface. The calculations were performed for bulk STO and LNO separately, and the alignment was obtained by determining the averaged electrostatic potential of STO and LNO slabs with respect to vacuum. Then the valence-band maximum (VBM) in bulk STO and the Fermi level in bulk LNO were determined with respect to the averaged electrostatic potential, and aligned to that in the bulk region of the slabs. The slabs were composed of 8 unit cells along of the *c* direction, and a vacuum layer of 12 Å. The calculations for STO were performed using the HSE06 hybrid functional whereas the calculations for LNO were performed using the Perdew-Burke-Ernzerho (PBE) generalized gradient approximation (GGA) for the exchange-correlation, as implemented in the VASP code[18]. The integrations over the Brillouin zone were replaced by sums of a mesh of 6x6x6 special k-points for STO and 12x12x1 for LNO, using 5-atom cubic unit cells. For the slabs, 6x6x1 k-points meshes were employed. A cutoff of 400 eV was employed for the plane-wave basis set. The value found was 2.25 eV, in good agreement with 1.75 eV and 2.25 eV from prior soft and hard x-ray measurements respectively, based on core-level binding energies as fixed references to the VBM[19].

As one further aspect of these calculations, we show in Fig. 2 the correctly offset band structures of bulk STO and bulk LNO, together with the projected DOS for LNO. Note the positions of the $t_{2g}$- and $e_g$-derived bands near E$_F$ and in the STO bandgap. This means that photoemission can see these bands in energy through the gap in STO, as pointed out previously in a similar study on LSMO/STO[13]. The projected densities of states also indicate that a peak due to $e_g$ is expected nearest E$_F$, and a peak due to $t_{2g}$ just below that.

**B. Electronic structure of the multilayer**: Density functional calculations were carried out for $(LNO)_3/(STO)_3$ superlattices closely related to our sample using the all-electron full potential linearized augmented plane wave method as implemented in the WIEN2k code[20]. Although the STO layers are thinner than the 4 u.c. of our sample, this is not expected to lead to significant changes in the behavior of the LNO or LNO/STO interface electronic structures, and we have checked this with trilayer calculations. Electronic correlations beyond the generalized gradient approximation of the exchange correlation[21] are considered within the LDA/GGA+U approach[22], with U = 4 eV and J = 0.7 eV for Ni and Ti $3d$ states, and U = 7 eV and J = 0 eV for the La $4f$ states. For further details on the calculations see Refs. 23 and 24. The results for a $(LNO)_3/(STO)_3$, superlattice are presented in Fig. 3. The relaxed structures shown in Figs. 3(a) reveal significant octahedral tilts in LaNiO$_3$, with these extending into SrTiO$_3$. The multilayer contains two polar interfaces of different type: an n-type [TiO$_2$/LaO(+)] and a p-type [SrO/NiO$_2$(-)]. The layer resolved DOS in Fig. 3(b) shows that the compensation of these polar interfaces takes place exclusively in the LNO part, while, unlike LaAlO$_3$/SrTiO$_3$, Ti $3d$ states do not participate and Ti remains 4+, as suggested also by the band alignment in Fig. 2. The main feature is the upward shift of Ni $3d$ bands from the n- towards the p-type interface, indicating a difference in band filling that is also reflected in the difference in magnetic moments (1.30/1.07 $\mu_B$ at the n/p-interfaces, respectively). For reference, the ideal magnetic moment of the Hund's Rule ground state of Ni$^{3+}$ $t_{2g}^6 e_g^1$ is 1.0 $\mu_B$. These results also indicate that the density of states near $E_F$, as averaged over about 1 eV near $E_F$, is lower for the n-type, which, by comparison with prior experiment[6] and current results to be discussed in Fig. 4, may suggest that it is somehow more prevalent in the multilayer growth.

Comparing the DOS results in Figs. 2(c) for bulk LNO and in Fig. 3(b) for the multilayers also shows a shift of the main features associated with LNO downward in energy in the multilayers by 0.6 eV for a p-type interface and about 1.4 eV for an n-type interface, leaving a much lower DOS near $E_F$ that is consistent with a prior standing-wave photoemission experiment[6], and a reanalysis of these prior results in the present study.

Finally, $3d$-projected densities of states for the multilayer in Fig. 3(a),(b) (not shown here) indicate that the relative $e_g$ character is strongest near $E_F$ and for the bottom bands near ~6 eV binding energy, in good agreement with Fig. 2(c) for bulk LNO.

These multilayer theoretical results are discussed in more detail elsewhere[24].

**IV. Results and discussion**

The samples were first characterized by angle-integrated standing-wave (SW) photoemission, through fitting experimental SW rocking curves (RCs) to x-ray optical calculations of photoemission for both core levels and the valence-band, following the method of Refs. 6, 13, and 14, with our recheck of the work in Ref. 6 being presented in Fig. 4. From this prior analysis[6] and our reconfirmation of it, it was found that the top STO layer had a somewhat smaller thickness of 9 Å (compared to the nominal as-grown 11.7 Å) and that the interface interdiffusion/roughness profiles were as described in Fig. 4, including an "insulating" layer of much reduced DOS for binding energies close to $E_F$ and next to the interface and a metallic central layer of LNO. An analysis of both soft and hard x-ray core-level intensities confirmed a carbon-containing layer on the STO surface with a thickness of about 9 Å if the density is assumed to be that of solid CO; the layer would be thinner if its density is higher. Previous TEM measurements confirmed the degree of intermixing at the interfaces shown in Fig. 4, and the top surface showed clear step-and-terrace structure as measured by AFM[7,25]. X-ray diffraction exhibited superlattice peaks up to third order as well as sharp Kiessig oscillations originating from the bottom and top of the stack, which shows that the sample maintains crystal registry throughout the entire sample thickness.

Core-level spectra of all of the constituent atoms, including carbon in a surface contaminant peak, were obtained with both soft-x-ray and hard x-ray excitation, and standing-wave RCs were measured for all of them with a more surface sensitive photon energy of 833.2 eV. A summary of these results is shown in Fig. 5, including the RCs for each peak in 5(a), the spectra for an angle maximizing the intensity from STO in 5(b), and a survey scan showing the regions from which these results were derived in 5(d). It is noteworthy in Fig. 5(a) that Ti $2p$, Ti $3s$ and Sr $3d$ all have essentially the same RC, as expected since they all are present in the STO layer. Likewise, La $4d$ and Ni $3p$ have essentially the same RCs, as both are present in the LNO layer. The C $1s$ RC is distinctly different from all of these, as it must represent the thin layer of C-containing surface contaminants on the sample. Beyond this, the single-peak Ti $2p$ spectra are not found to change over the full RC, and not to show any indication of a feature at lower binding energy by about 1.0-1.5 eV that has been seen in prior work and taken to either represent a $Ti^{3+}$ state[26] and/or the influence of surface oxygen vacancies on Ti[27,28]. Our results thus permit concluding that the Ti in the multilayer is predominantly in the 4+ state, as expected for STO, with very little or no indication of $Ti^{3+}$ or the influence of surface oxygen vacancies. This conclusion is supported by deeper penetrating hard x-ray photoemission data at several photon energies (2550, 3237.5, 4000, and 5953.4 eV), with discussion of these results appearing elsewhere[16].

It is interesting to now look at both soft and hard x-ray photoemission data that are in the MEWDOS limit, and some examples of these are shown in Fig. 6. Hard x-ray photoemission results are shown for two take-off angles $\theta_e$: more bulk sensitive at 88° and more surface sensitive at 45°, and the inherently more surface sensitive soft x-ray results with takeoff angles averaged over 55°-95° (i.e., from slightly past normal toward the surface), with data obtained in an angle-resolving detection mode. The soft x-ray results represent an integral over the two angles in a SWARPES measurement corresponding to $k_x$ and $k_y$ as measured by accumulating detector images in binding energy and $k_x$ at different tilt angles $\beta$ corresponding to $k_y$ (see Fig. 1). Five special energy regions are noted in this figure:

- region **1** that from Figs. 2 and 3 and 3$d$-projected DOSs for the multilayer[24] we expect to be predominantly states derived from Ni $e_g$,
- region **2** to be from Ni $t_{2g}$,
- region **3** to be from the Bottom of the bands due to a mixture of the lowest-lying bands of STO and LNO,
- features **4** and **5** due to a more complex origin from both STO and LNO.

We can unambiguously assign regions **1** and **2** to the aforementioned Ni 3$d$ states due to the absence of spectral weight of STO in this energy range, by analogy with a prior study of STO on LSMO[13], and as more quantitatively determined from previous band offset measurements on the same sample, as illustrated in Figs. 2 and 3. The intensity in region **3** is more difficult to assign unambiguously, at the theoretical calculations in Figs. 2 and 3 suggest that LNO should be the dominant contributor, but RCs of the valence-band features to be presented in Fig. 11 suggest more STO contribution. Features **4** and **5** over ~2-7 eV binding energy we expect to be a more complex mixture of states from STO and LNO, as expected from Figs. 2 and 3. Our first-principles calculations also show that the Fermi level in LNO, which crosses the Ni $e_g$ bands, is about 2 eV higher than the VBM of STO which is mainly of O 2$p$ character, as shown in Fig. 2. The Ni $t_{2g}$ states are for this case more pronounced compared to the $e_g$ states, which is due to some combination of inherently higher densities of states, as shown in Fig. 2(c), and matrix-element effects. Taken together, these results permit us to cleanly identify two regions **1** and **2** that should be primarily associated with LNO, with 3 a likely mixture of the two, but possibly with more LNO, and we will focus on these three in discussing the SWARPES results.

The experimental geometry of the SWARPES measurements on our [4 u.c. LNO/3 u.c. STO]$_{10}$ sample as scanned in $k$ space is summarized in Fig. 7. In Fig. 7(a), the method of accumulating the data via a mechanical scan of the tilt angle $\beta$ is shown (cf. Fig. 1). In Fig. 7(b), the region in $k$-space spanned by the detector, which has an angular range of ~ ±20°, is indicated, with the red arc representing the locus of points measured. In arriving at this figure, we assume excitation from a valence electronic state with momentum $k_i$ and a free-electron final state with $k_f$ according to the usual direct transition selection rule of ARPES: $k_f = k_i + g + k_{hv}$, where the magnitude of $k_f$ inside the sample for 833.2 eV excitation is about 14.87 Å$^{-1}$, $g$ is the appropriate reciprocal lattice vector supporting the direct transition, $k_{hv}$ is the photon wave vector with magnitude of 0.422 Å$^{-1}$ and must be allowed for due to effects beyond the dipole approximation. The initial state wave vector involved in a given transition is thus $k_i = k_f - k_{hv} - g$ , and we illustrate in Fig. 7(b) the application of this equation for a special transition in which $k_f - k_{hv}$ lies along the [001] direction, for which the relevant $g$ has a magnitude of 9(2π/a) = 9(1.61) Å$^{-1}$ = 14.49 Å$^{-1}$. For this case, with the x-rays incident at 60° with respect to the electron emission direction, the angle between $k_f$ and $k_f - k_{hv}$ is 1.21°. The incidence angle relative to the surface was varied over five values, whose choice we now discuss, with two-dimensional $k_x$ -$k_y$ ARPES patterns being measured at each angle for all valence binding energies.

To provide further insight into the nature of the surface in $k$-space of our measurements, Fig. 8 shows several three-dimensional images of the ARPES results. These data have been derived from the raw patterns via a consecutive correction for the effects of density-of-states like intensity modulation in energy, and photoelectron diffraction modulation along the $k_x$ and $k_y$ angular directions, as discussed in detail elsewhere by Gray et al.[13].

To now provide a more quantitative idea of the SW approach that we will apply in ARPES, Fig. 9(a) shows RCs of Ni 3$p$ and Ti 2$p$ core levels, which have been measured for our [4 u.c. LNO/3 u.c. STO]$_{10}$ sample by scanning the x-ray incidence angle through the multilayer Bragg angle, as introduced before with Fig. 5. In such an RC, the SW moves vertically by one half of its period, with the period in turn being very close to the multilayer period of $d_{ML}$. These RCs are thus representative of the depth-dependent photoemission in the two different layers of the multilayer, and the experimental points in Fig. 9(a) are compared with x-ray optical calculations of the photoemission intensity from Ni 3*p* and Ti 2*p*, making use of the specially-written YXRO program described in detail elsewhere[15]. The agreement between experiment and theory is excellent, and the dramatic difference in the shape and phase of the two curves with respect to one another are a consequence of the SW sweeping through the sample as incidence angle is changed. The left panels of Figs. 9(b) and 9(c) further show the calculated depth dependent intensities of Ti 2*p* and Ni 3*p* intensity in each of the first three layers of the sample for the two specifically chosen incidence angles, together with C 1*s* intensity from the contaminant overlayer that has been observed previously[6] and discussed above. The right panels of Figs. 9(b) and 9(c) further show detailed depth-resolved intensity profiles as a function of x-ray incidence angle, as well as vertical line cuts, in 9(b) for the pair of angles 2 and 3 that show the largest difference in the two RCs of the five angles studied, and in 9(c) for the pair 4 and 5 showing the smallest difference. Each line cut here corresponds to a depth-selective measurement of photoemission, which is modulated by the SW electric field profile and the exponential decay of intensity according to the IMFP (or rather the effective attenuation length allowing for elastic scattering) of the photoelectrons. As can be seen, at each incidence angle there are unique points at which the intensity is magnified and suppressed and these variations can in principle be used to extract depth sensitive information by taking appropriate differences of data obtained at different angles. More specifically, angle 2 (15.7°) represents nearly maximizing the Ti 2p signal from the top STO layer and nearly minimizing the Ni 3p signal from the underlying LNO layer, thus enhancing the intensity from the interface between the two. Angle 3 (14.9°) represents nearly minimizing the Ti signal from the top STO layer and nearly maximizing the Ni signal from the underlying LNO layer, thus enhancing the intensity from the "bulk" of the LNO layer, as illustrated by the x-ray optical calculations in the right panel of Fig. 9(b). Taking the difference of these two angles thus should yield SWARPES patterns that are more characteristic of the interface. From Fig. 9(a) ca, it is true that the difference between 15.9° and 14.6° might have more cleanly focused on the interface, as these represent more accurately the extrema of the Ti *2p* and Ni *3p* intensities, but these positions were not known accurately while acquiring the data.

From this figure, we thus expect that a difference of data obtained and angles 2 and 3 should provide the most sensitive information on the properties of the STO/LNO interface, as discussed previously for LSMO/STO by Gray et al.[12,13] (see especially Figures S3, S4, and S4 in the Supplementary Information of this paper), and that a difference of data obtained at angles 4 and 5 should be much less sensitive. Looking ahead to future studies of this type, we suggest fully analyzing the core rocking curves during the experiment before choosing the optimum angles for further SWARPES measurements, thus even more fully optimizing the selectivity to the interface.

We now focus on our low-temperature SWARPES results, corrected for both phonon-induced DOS intensity and photoelectron diffraction with the method of Gray et al.[13], and integrated over three energy ranges of **1**-$e_g$: 0.14-0.46 eV, **2**-$t_{2g}$: 0.74-1.36 eV, and **3**-Bottom: 7.35-8.05 eV. In. Fig. 10, we compare ARPES images taken at angles 2 and 3, and the difference 3 - 2, as well as images at angles 4 and 5, and the difference 5 – 4. Intensity scales for the ARPES images and their differences are also shown to provide an idea of the strength of the state dispersions seen. For all of the Bottom ARPES images, there is a clear square grid pattern that is very similar to that seen in a prior study of LSMO/STO[13], to be discussed further below. The Ni 3$d$ $t_{2g}$ and Ni 3$d$ $e_g$ ARPES images show much more complex dispersion patterns that are again very similar to those of LSMO/STO. The patterns are also very similar for all of the angles 2, 3, 4, and 5, as expected. The difference plots by contrast are quite different between 3 – 2 and 5 – 4, with 3 – 2 exhibiting much more structure, as expected from the discussion of Fig. 9. In fact, there is no reliable structure seen in the 5 – 4 difference data. These results are in full agreement with similar trends seem before in LSMO/STO[13], and confirm the ability of SWARPES to isolate differences in electronic structure associated with buried interfaces.

The Ni 3$d$-derived images can also be compared to the previous standing-wave XPS work of Kaiser et al.[6] as follows. If these maps are integrated over emission angle to obtain a single intensity for every incidence angle, the results should to a good approximation agree with the RCs obtained for the Ni 3$d$ states in this prior work, carried out at room temperature and thus in the MEWDOS limit. Fig. 11 compares several angle-integrated ARPES intensities: Ni 3$d$ $t_{2g}$, Ni 3$d$ $e_g$, and Bottom, as well as the core-level Ni 3$p$ variation, together with a theoretical curve calculated for the sample geometry of Fig. 4 using our x-ray optics program[15]. Although less densely measured in angle than Ni 3$p$, the statistical scatter in the Ni 3$d$ $t_{2g}$ and Ni 3$d$ $e_g$, points is very small, they agree very well with one another over the entire angle range, as described by the dashed orange curve drawn through the points, and they unambiguously show a RC shift of about 0.2° relative to the Ni *3p* RC. In a prior analysis of similar RC results for LNO/STO based on MEWDOS data, this shift has been interpreted as a suppression of these states near the Fermi level[6]. So this same effect is seen in angle-integrated ARPES, confirming this earlier conclusion. The fact that the data for the Bottom states shows a very different profile from the Ni-derived states is also significant, as it in fact looks more like the Ti *2p* RC in Figs. 5(a) and 7, although much flatter in its variation. These states we attribute to a mixture of states from STO and LNO and their RC should thus be a weighted average of a STO and LNO characteristic-rocking curve. This explains the lowering of the amplitude of this RC, but its shape nonetheless suggests a greater weight of STO states, perhaps due to smaller attenuation of the photoemission signal from the topmost STO layer. This is the opposite conclusion from that reached by considering the theoretical results in Figs. 1 and 2, so the assignment of relative weights cannot be made conclusively.

Finally, in Fig. 12, we compare our SWARPES results for LNO/STO with those from a different, although closely related perovskite-based superlattice sample of $La_{0.7}Sr_{0.3}MnO_3/SrTiO_3$ (LSMO/STO). In both figures, alignment lines and circles to permit more easily seeing the similarities between the two sets of data are shown. Despite the fact that the patterns seen in the LNO/STO sample are due to Ni *3d* states, and the patterns seen in LSMO/STO sample are due to Mn *3d* states, the two both involve $e_g$ and $t_{2g}$ derived bands that can be seen through the STO gap, and in fact exhibit remarkably similar behavior. Comparing Fig. 2 here with Fig. S6 in Ref. 12, which compares the LSMO and STO band structures provides some confirmation that this should be true.

Beyond this, region **3** = Bottom bands in both systems again show the same simple-square pattern, which has also been predicted theoretically for LSMO/STO using both a free-electron final-state model and a much more accurate one-step model of photoemission[13]. Despite the different band filling of $Ni^{3+}$ ($3d^7$) in LNO, and Mn in $La_{0.7}Sr_{0.3}MnO_3$, which is, in the simplest picture, a mix of 2/3 $Mn^{+3}$ $3d^4$ and 1/3 $Mn^{+4}$ $3d^3$, the electronic configuration considering the crystal-field splitting $Ni^{3+}$ $t_{2g}^6$ $e_g^1$ and $Mn^{+3}$ $t_{2g}^3$ $e_g^1$ is expected to be very similar in character, due to the equivalence of electrons and holes in the configurations involved. The similarity of the overall ARPES patterns can thus be attributed to the dominant transition-metal configurations involved, the overall structural similarity of the LNO and LSMO crystal structures, and the fact that, in both cases, these Bottom states are very similar mixtures of metal *3d* and O *2p* that are not expected to depend strongly on the form of the states at much lower binding energies. Some key differences between the two systems are in the amplitude of the dispersive modulations, which are more pronounced in LSMO/STO in all panels. This we attribute to the less ideal interfaces in LNO/STO, with an estimated roughness of one u.c.[7]

### V. Conclusions

We have in this study demonstrated several aspects of using the additional depth selectivity of soft x-ray standing-wave photoemission, or more particularly, standing-wave angle-resolved photoemission (SWARPES), for studying the electronic structure in buried layers and interfaces, including momentum-resolved electronic structure. The experimental data are also augmented by state-of-the-art density functional theory calculations for the electronic structure of both the bulk parent materials LNO and STO, and the multilayer. This study thus follows the pioneering work of Gray et al. for LSMO/STO[13], and represents a third system, together with GTO/STO[14], that has been studied with ARPES making use of standing-wave and x-ray optical effects. We have studied a [LaNiO(3 unit cell)/SrTiO$_3$(3 unit cell)]$_{10}$ multilayer with resonant soft x-ray excitation at 833.2 eV to enhance reflectivity and thus SW strength, coupled with complementary characterization using more depth sensitive hard x-ray excitation at several energies. Core-level intensities are monitored through rocking curves (RCs) and together with x-ray optical modeling used to determine those soft x-ray incidence angles at which different depths in the sample are probed, including a pair of angles expected to be the most sensitive to the LNO/STO interface. The RCs permit confirming the depletion of the density of states near the Fermi energy associated with a metal-insulator transition. The SWARPES results for the electronic states derived from Ni $e_g$ and $t_{2g}$, as well as states at the Bottom of the valence bands, are in fact found to show very similar dispersions to those seen in LSMO/STO[13], but in general with smaller differences when using different incidence angles to enhance the LNO/STO interface. This difference between the two systems we attribute to the greater difficulty of growing LNO/STO, which can lead to rougher, less well-defined interfaces. But the fact that SWARPES can, even for this case with only 10 multi-layer repeats, still provide strong SW effects and interface-specific densities of states and momentum-resolved information is encouraging for its future applications to other multilayer systems.

**Acknowledgements**

Primary support for this work is from the MURI program of the Army Research Office (Grant No. W911-NF-09-1-0398). The Advanced Light Source, A.B., W.C.S., and C.S.F. are supported by the Director, Office of Science, Office of Basic Energy Sciences, Materials Sciences and Engineering Division, of the U.S. Department of Energy under Contracts No. DEAC02-05CH11231 at the Lawrence Berkeley National Laboratory, and No. DE-SC0014697 at the University of California Davis. Additional support for C.S.F. has come from the Laboratory Directod Research Development Program at LBNL. During the final writing of this paper, CSF received salary support from DOE Grant DE-SC0014697 through the University of California Davis. The HXPS measurements at BL15XU of SPring-8 were performed under the approval of NIMS Beamline Station (Proposal No. 2011A4606). S.U. and K.K are grateful to HiSOR, Hiroshima University and JAEA/SPring-8 for the development of HAXPES at BL15XU. The Advanced Light Source is supported by the Director, Office of Science, Office of Basic Energy Sciences, of the U.S. Department of Energy under Contract No. DE-AC02-05CH11231. A.J. and C.G.V.d.W. were supported by the US Army Research Office (W911-NF-11-1-0232). Computational resources were provided by the Extreme Science and Engineering Discovery Environment (XSEDE), supported by NSF (ACI-1053575). S.N. received support in the completion of this work from the Jülich Research Center. G.K.P. also thanks the Swedish Research Council for financial support. A.R. was funded by the Royal Thai Government and C.C. was funded by GAANN program through UC Davis Physics Department. C.S.F. has also been supported during the writing of this paper for salary by the Director, Office of Science, Office of Basic Energy Sciences, Materials Sciences and Engineering Division, of the U.S. Department of Energy under Contract No. DE-AC02-05CH11231, by the Laboratory Directed Research and Development Program of Lawrence Berkeley National Laboratory under the same contract, and by the LabEx PALM program Investissements d'Avenir overseen by the French National Research Agency (ANR) (reference: ANR-10-LABX-0039). R.P. and A.B.R. acknowledge funding by the German Science Foundation, SFB/TR80 (project C3 and G3) and BaCaTeC.

**Figure 1**

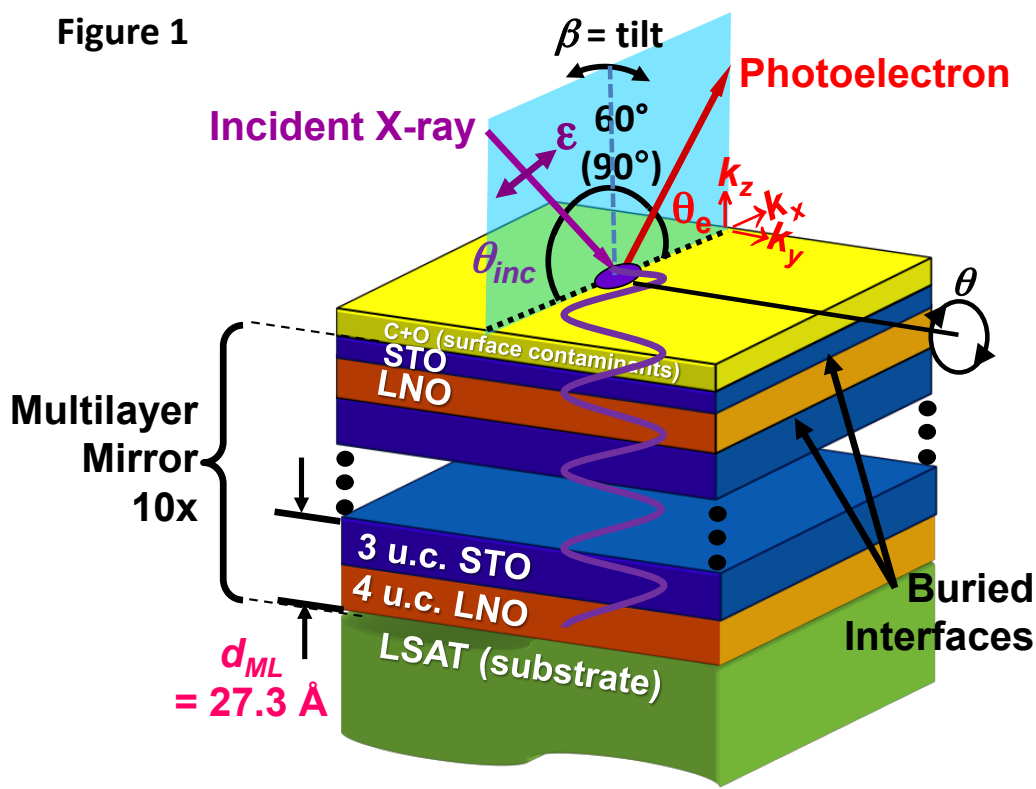

Figure 1. Schematic of the LaNiO$_3$/SrTiO$_3$ (LNO/STO) superlattice sample structure and experimental geometry: A multilayer mirror consisting of 10 bilayers (4 unit cells of LaNiO$_3$ /3 unit cells of SrTiO$_3$ ) with a period of 27.3 Å, except for the top bilayer, for which the thickness of the top STO layer is found to be reduced such that the bilayer period is 24.6 Å (ref. 6). The sample was rotated on two axes, $\theta$ and $\beta$ = tilt. The geometry for the soft x-ray standing-wave ARPES measurements was 60° between incidence and exit and for the hard x-ray photoemission measurements 90°. In ARPES, all three angles were varied: the x-ray incidence angle $\theta_{inc}$, the coupled electron emission angle $\theta_e$ and the sample tilt angle $\beta$. The radiation is p-polarized, as indicated by the vector $\varepsilon$. Also shown is a schematic standing wave intensity profile, with period equal to the bilayer distance $d_{ML}$.

**Figure 2**

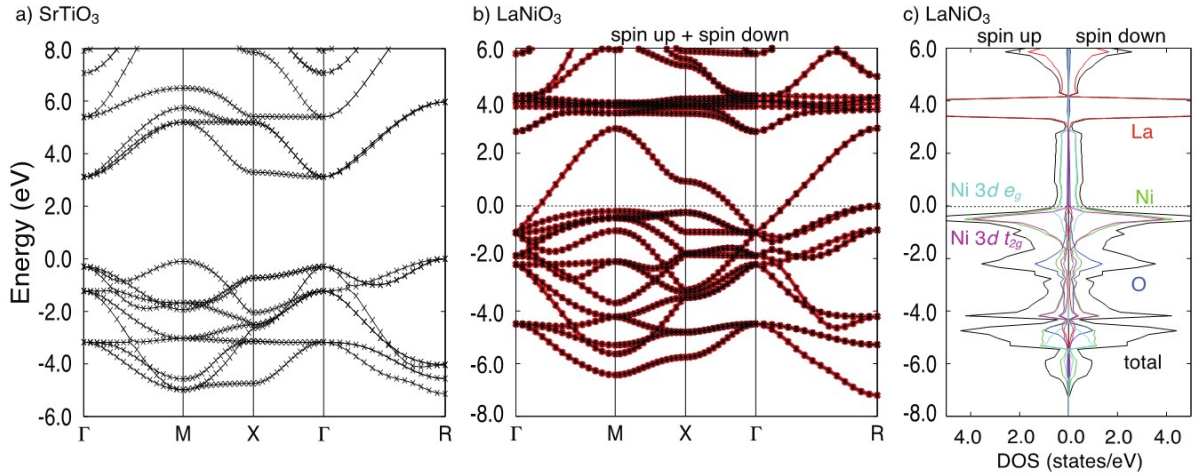

Figure 2. The electronic structures of bulk STO and LNO. (a),(b) The calculated band structures of (a) the band insulator STO and (b) metallic LNO. The calculation for STO was performed using the HSE06 hybrid functional, and that of LNO using the GGA PBE functional. The band offset Fermi level in LNO with respect to the valence band of STO (~2.2 eV) was calculated using the HSE06 hybrid functional as in Ref. [17]. (c) The atom and d-orbital resolved densities of states derived from (b).

## Figure 3

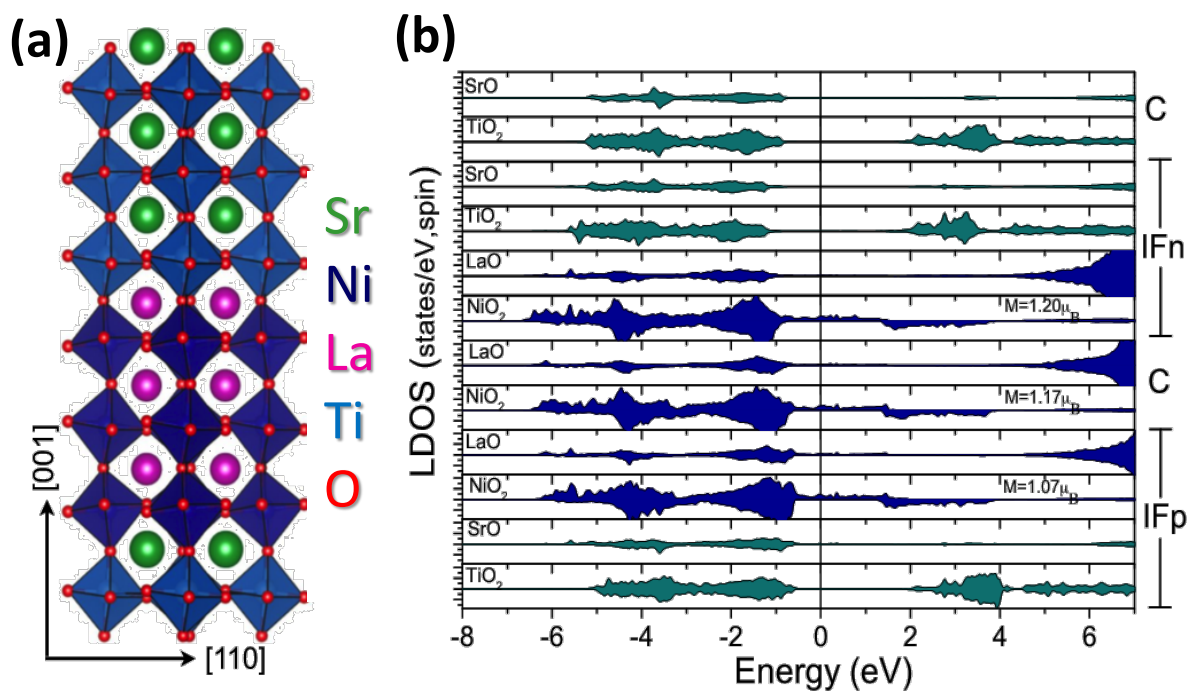

**Figure 3** Spin-polarized density functional calculations (DFT+U) for an LNO/STO multilayer. (a) The atomic configuration for an LNO-3 u.c/STO-3 u.c. multilayer, with the two types of ideal interfaces indicated: an n-type (IFn) [TiO$_2$/LaO$_2$(+)] and a p-type (IFp) [SrO/NiO$_2$(-)]. (b) The layer-resolved total densities of states indicating: the net magnetic moments on Ni in the interface and middle layers of LNO and the bond length of La-Sr (d$_{La-Sr}$) across each interface. From ref. 24.

**Figure 4**

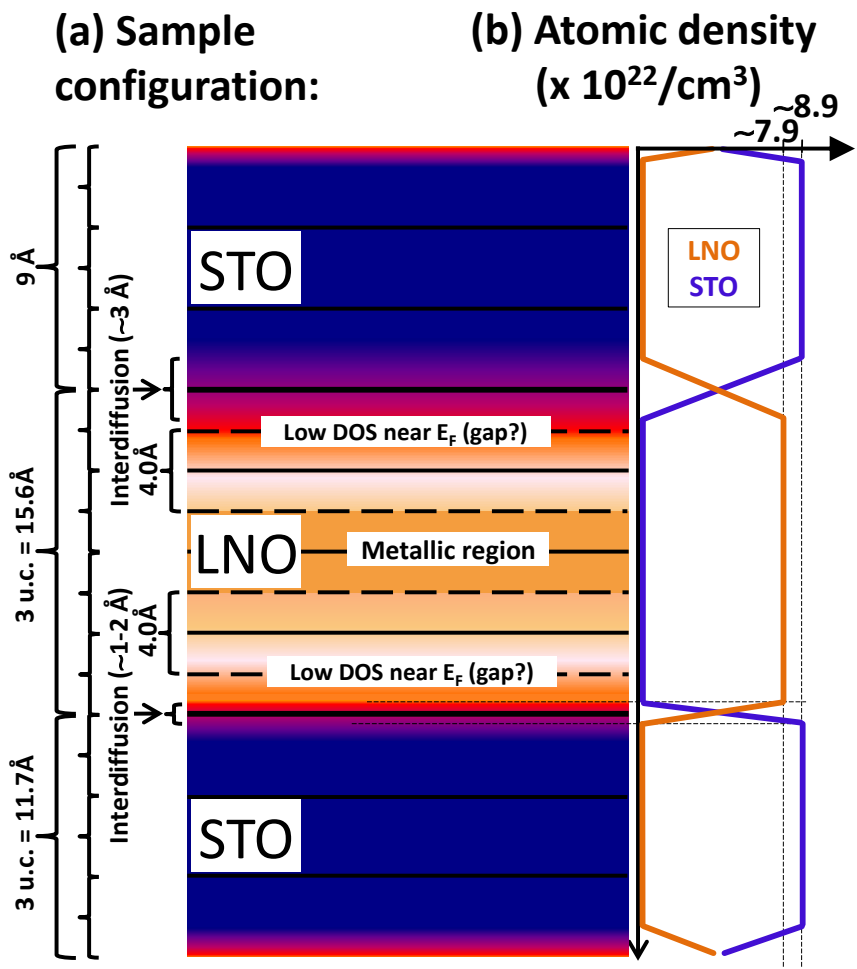

**Figure 4.** (a) The schematic depth profile of the top trilayer in our sample, as derived from an analysis of core-level and valence-band rocking curves, and confirmed by a reanalysis of the results of ref. 6. Metallic and semiconducting/insulating portions of LNO with approximate concentration gradients are indicated. (b) The atomic number density of LNO (orange) and STO (indigo) corresponding to each layer, estimated from their bulk densities.

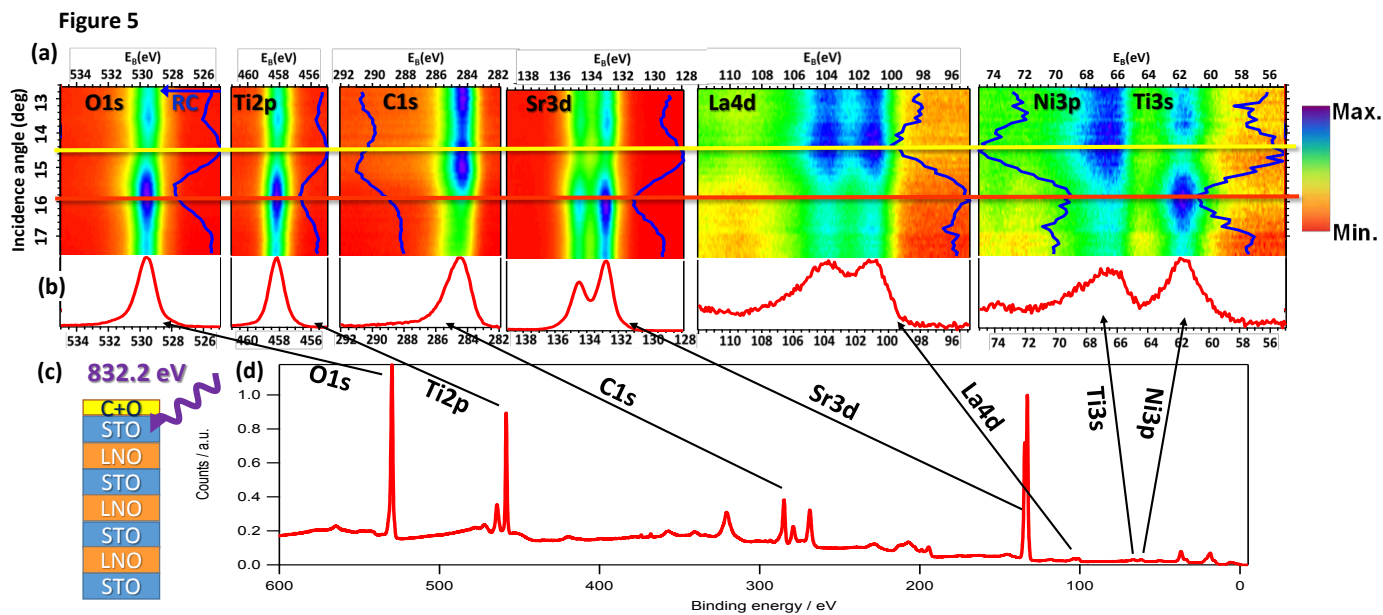

**Figure 5.** A summary of core-level standing-wave measurements at 833.2 eV. (a) Color-scale intensity plots in incidence angle versus binding energy of individual spectra from all atoms present in the sample. The blue curves are rocking curves (RCs) based on intensities derived from peak fitting each spectrum. (b) The spectral forms for each core-level for an angle of incidence maximizing the signal from STO, as judged from the RCs of Ti 2p and Ti 3s (red reference line). Note that this is also a maximum for Sr 3d, as expected. The yellow reference line represents an angle for which the LNO signal is maximized. (c) The sample configuration. (d) A survey spectrum indicating the region from which each spectrum is derived.

Figure 6

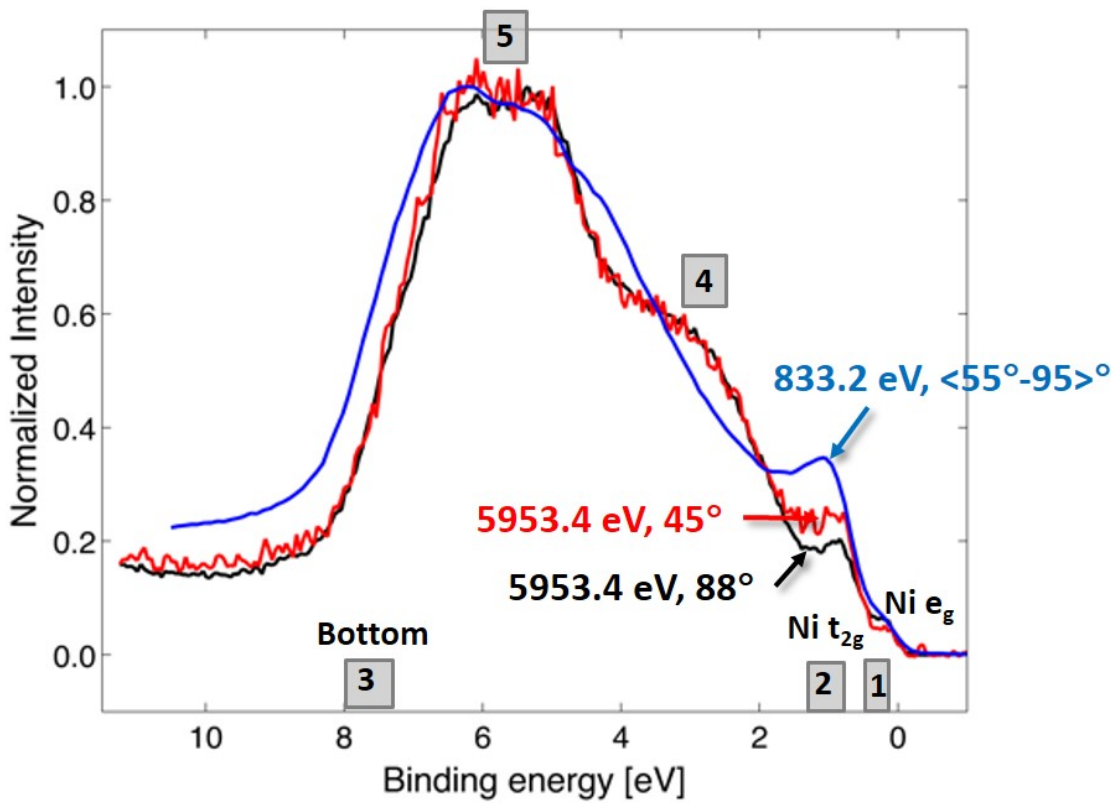

Figure 6. Valence-band hard and soft x-ray photoemission in the matrix-element-weighted density-of-states (MEWDOS) limit from the 4 u.c. LNO/3 u.c. STO multilayer. Five energy regions of 1 ≈ Ni $e_g$, 2 ≈ Ni $t_{2g}$, 3 = bottom of the LNO valence band, and 4,5 mixtures of LNO and STO bands are indicated. The hard x-ray emission angle $\theta_e$ was varied from 88° near normal (more bulk sensitive) to 45° off normal (more surface sensitive), suggesting that $t_{2g}$ may be more enhanced near the surface of the sample. The soft x-ray results were averaged from ARPES over 55°–75° and again show the same features, but with different relative intensities due to matrix-element effects.

Figure 7

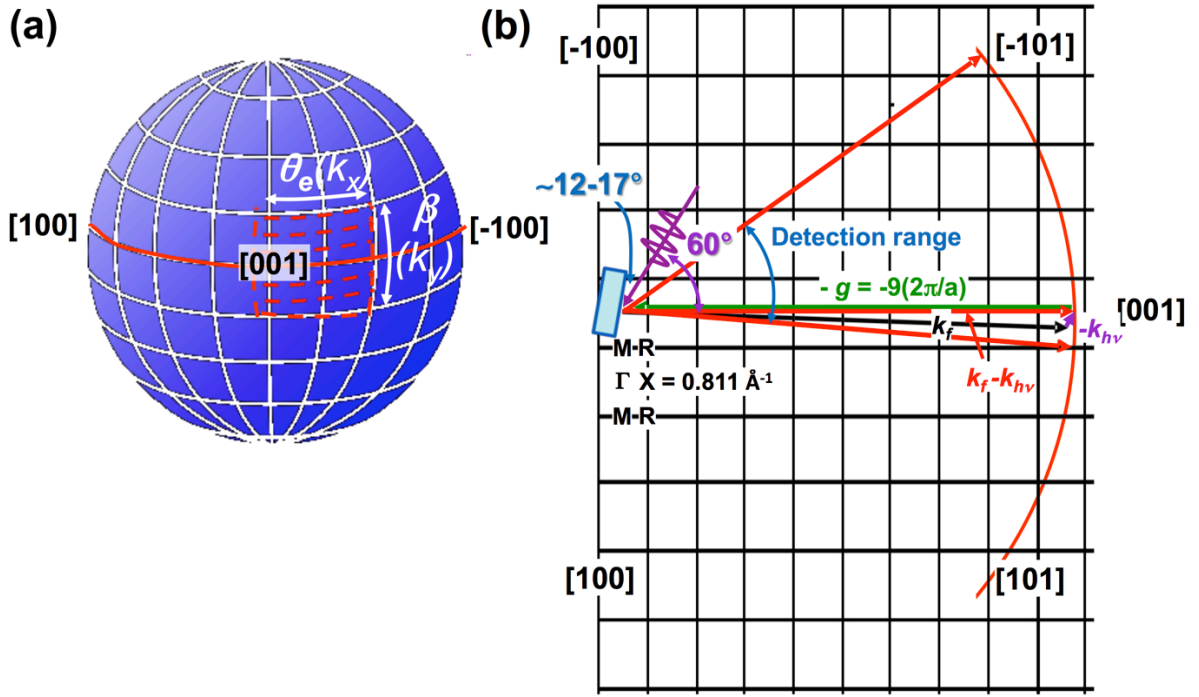

Figure 7. The experimental ARPES configuration in *k*-space, for 833.2 eV excitation. (a) The section of a sphere in $k_f$-$k_{hv}$ measured in ARPES through the two angles $\theta_e$ and $\beta$ (dashed lines). (b) A cross section through this sphere in the [100]-[001]-[-100] plane, showing various vectors and angles involved, including the special case of electron detection for $k_f$-$k_{hv}$ along [001].

**Figure 8**

**(a)-1, $e_g$**

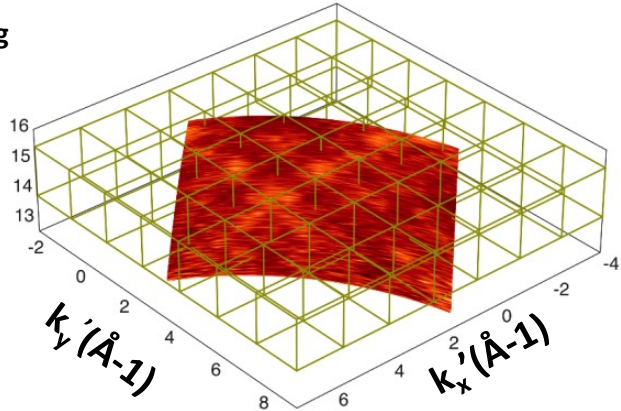
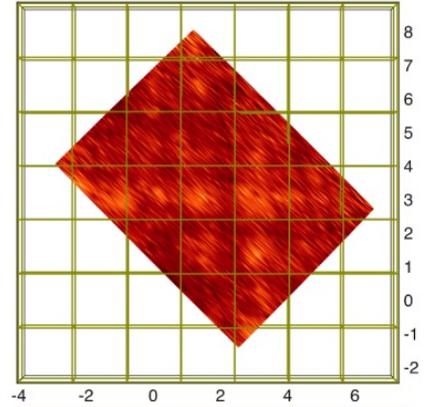

**(b)-2, $t_{2g}$**

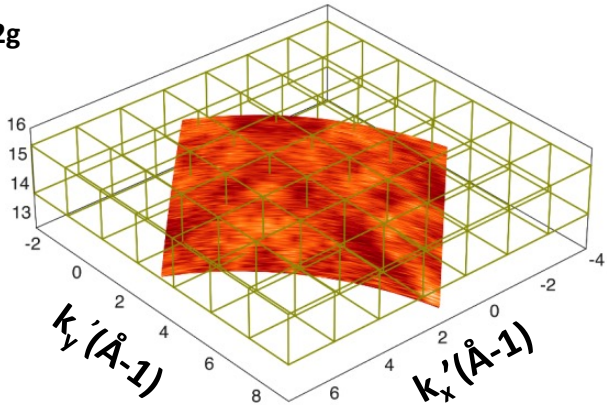
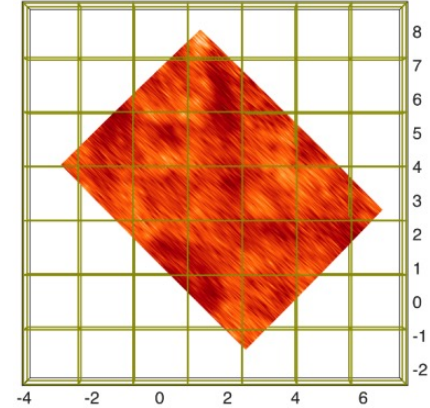

**(c)-3, Bottom**

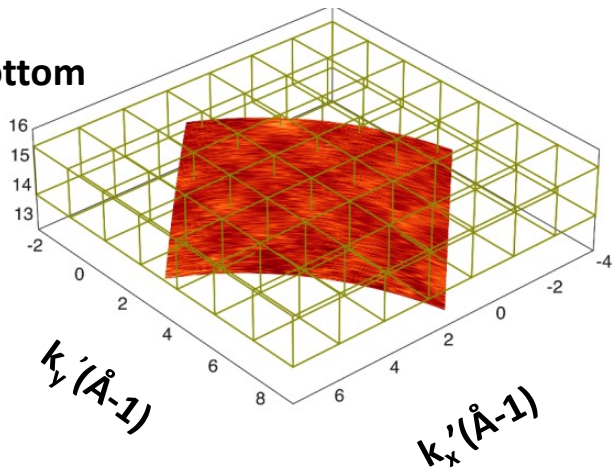
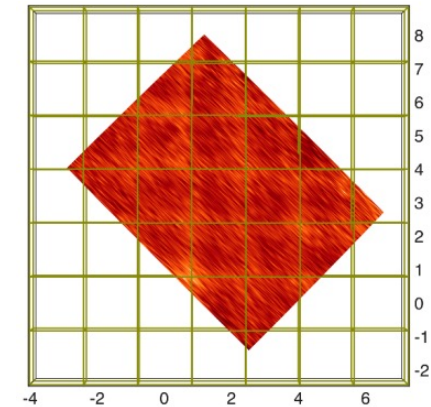

**Figure 8. Three-dimensional plots of the ARPES images at 833.2 eV excitation for the three valence energy regions corresponding to (a) 1-$e_g$, (b) 2-$t_{2g}$, and (c) 3-Bottom. The left panels show a perpective view, and the right panels a view in along the $k_z$ axis. Here, $k_x'$ and $k_y'$ are along the crystal x and y axes, and rotated by *45°* around $k_z$ from $k_x$ and $k_y$ as defined in Figs. 1 and 6.**

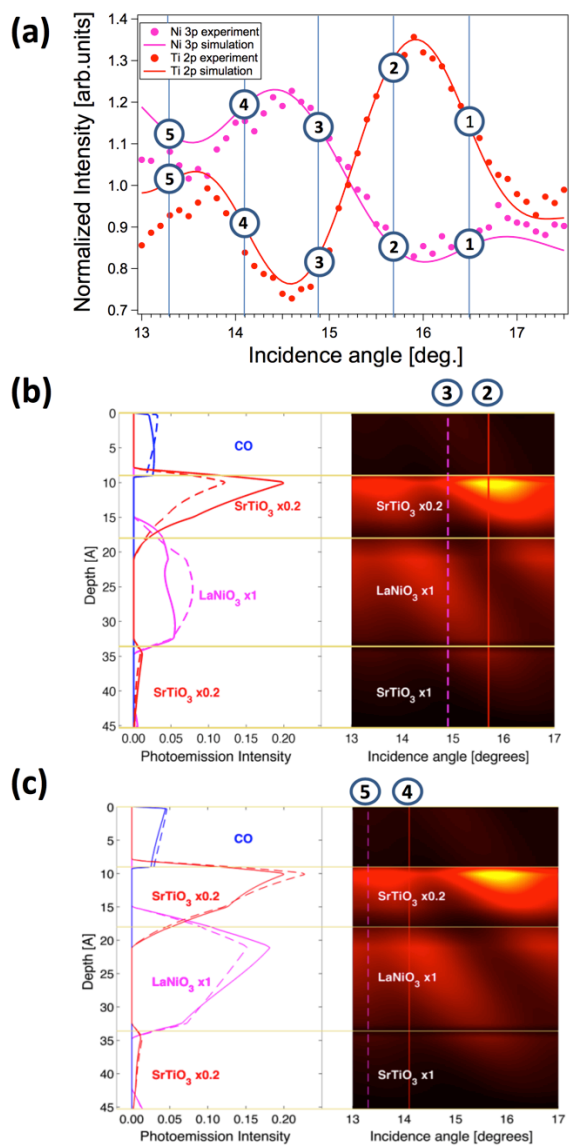

**Figure 9**

Figure 9. (a) The Ni 3*p* and the Ti 2*p* rocking curves, as measured (points) and calculated (curves), including an indication of the 5 different angles at which ARPES measurements were taken. (b) and (c) The simulated photoemission x-ray intensity profiles over the angular range of the Ti 2*p* and Ni 3*p* rocking curves (right panels) including line cuts of the profiles at the angles 2 and 3 (b) and 4 and 5 (c) indicated. By taking differences between 3 and 2 in (b) one expects to maximize the photoemission from the center of the LNO layer, whereas the photoemission profiles of 4 and 5 in (c) look quite similar and little difference is expected.

Figure 10

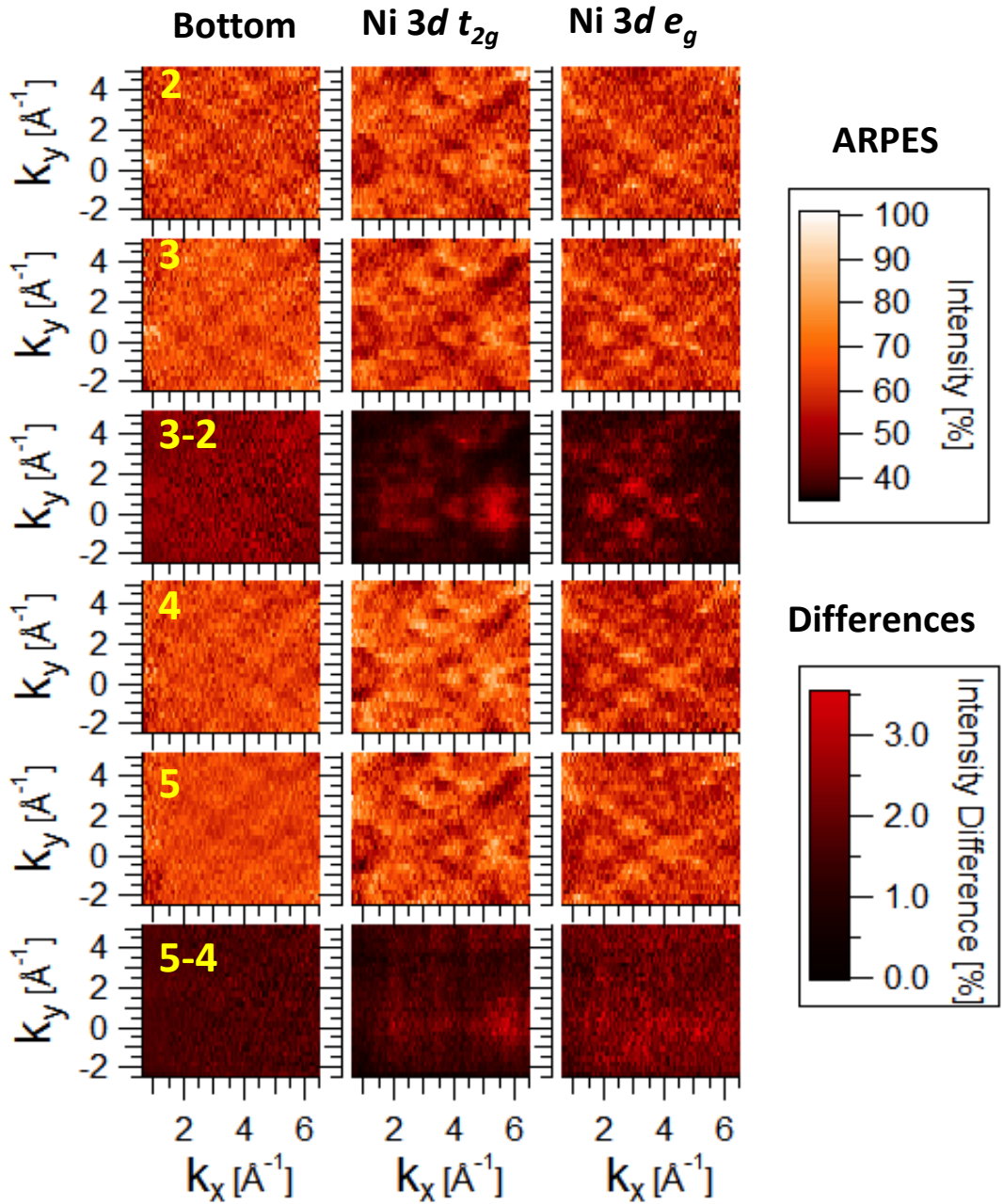

**Figure 10.** ARPES $k_x$-$k_y$ images with 833.2 eV excitation from the three binding energy regions corresponding to the bottom of the valence bands, the Ni 3d $t_{2g}$ and the Ni 3d $e_g$ obtained at angular positions 2 and 3 (cf. Fig. 9(a)), together with the difference 3-2, and also at angular positions 4 and 5 and the difference 5-4.

**Figure 11**

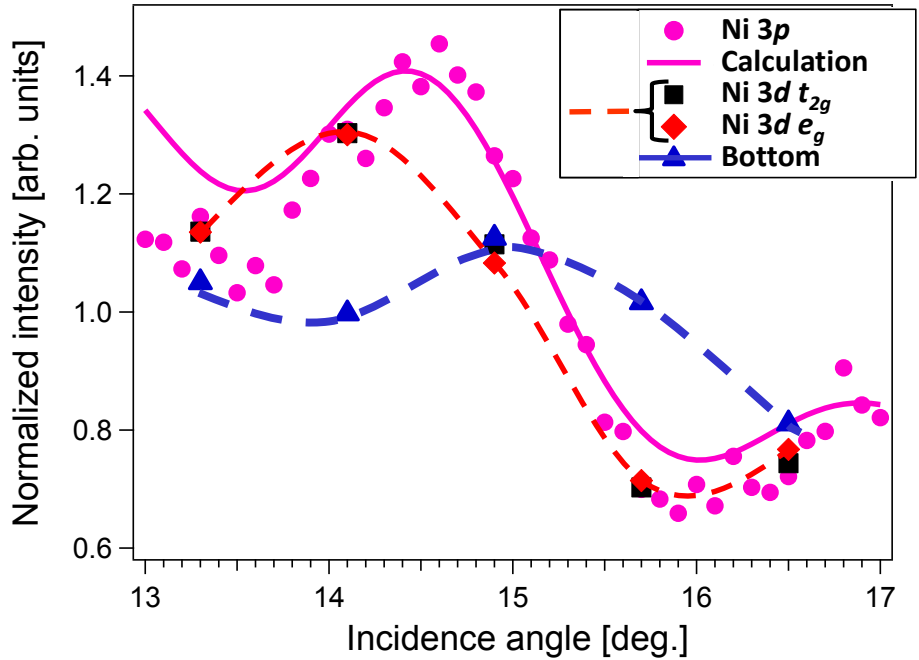

**Figure 11. The integrated intensities of Ni 3d $t_{2g}$ (black squares), Ni 3d $e_g$ (red diamonds), and the band bottom (blue triangles) are compared to the core-level Ni 3p rocking curve (experiment – magenta circles, simulation – magenta line). The curves for Ni 3d and Bottom are only guides to the eye. The horizontal shift of the integrated valence intensities relative to Ni 3p is consistent with the shift in thin-film LNO rocking curves previously published and believed to be linked to a depletion of Ni 3d density of states near the Fermi level at the LNO/STO interface [6].**

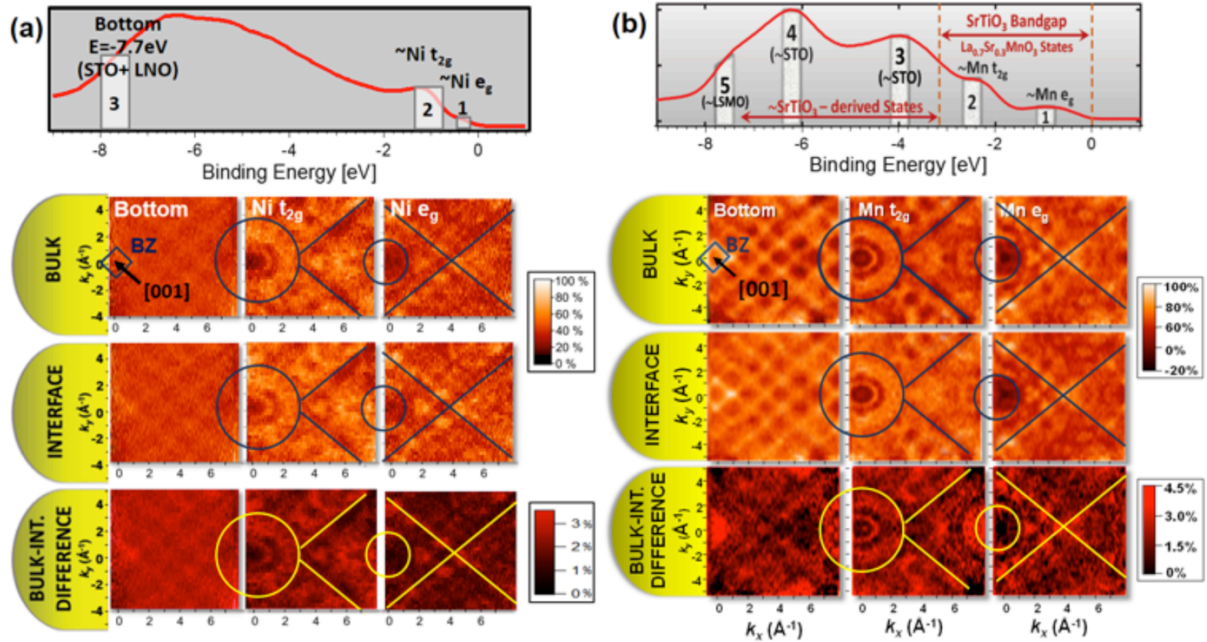

Figure 12. Comparison between (a) the current SWARPES results for LNO/STO and (b) those from a prior study of LSMO/STO by Gray et al. (ref. 12), with similar representative features highlighted in both panels by lines and circles. The photon energy in both cases is 833.2 eV. Irregularities at the edge of the detector in our data have led to cropping the LNO/STO maps slightly at the bottom as compared to the earlier work. The similarity between features is evident. However, the bulk-interface differences are much more pronounced in LSMO/STO. For the Bottom state, the difference for LNO/STO looks very much like the interface and bulk data, but for LSMO/STO is rich in structure.